# Sonic and Alfvénic Critical Point and the Solar Probe Mission

W. H. Matthaeus

*presentation to the Solar Probe Science and Technology Definition Team*

September 17, 2004

**Abstract**

Are critical points important in the Solar Probe Mission? This is a brief discussion of the nature of critical points in solar wind models, what this means physically in the "real" solar wind, and what can be expected along a nominal Solar Probe Orbit. The conclusion is that the regions where the wind becomes transonic and trans-Alfvénic, which may be irregular and varying, may reveal interesting physics, but the mathematically defined critical points themselves are of less importance.

**Critical Point in a Simple Model**

At a *critical point* in a system of ODEs, the solution is no longer determined uniquely, and one must appeal to boundary conditions or other physical effects outside the local mathematics to determine what actually will occur. Two such points are frequently discussed in coronal and solar wind physics– the *sonic* critical point and the *Alfvénic* critical point. The critical points appear when the coefficient of a derivative vanishes, which in the case the the two subject critical points, occurs because a wave speed, either the sound speed of the Alfvén speed, become equal to the flow speed. The mathematics of critical points emphasizes how the ambiguity of the solution is classified, understood and resolved. But from a physical point of view, including its relevance to the solar wind, the importance of these points lies in the implications that critical points have for the behavior of fluctuations and energy budgets.

The nature of the sonic critical point is easily understood using a very simple model – a spherically symmetric, steady, isothermal solar wind, consisting of equal number densities $n$ of protons (mass $m_p$) and electrons (mass $m_e$), and total mass density $\rho = nm$, and $m = m_p + m_e$. The relevant equations are (1) mass conservation ($\nabla \cdot \rho \mathbf{u} = 0$ for a radial outflow,)

$$\frac{1}{r^2}\frac{d}{dr}(r^2 \rho u) = 0, \qquad (1)$$

which implies that a mass flux $I = 4\pi \rho u r^2$, proportional to mass flux, is independent of $r$, and (2) the invisicid momentum conservation with a gravitational potential,

$$\rho u \frac{du}{dr} = -\frac{dP}{dr} - \rho \frac{GM}{r^2} \qquad (2)$$

where $M$ is the solar mass and $P = 2nkT \equiv \rho a^2$ for equal number densities $N$ and temperatures $T$ for protons and electrons. For the isothermal model, the sound speed is $a = \sqrt{2kT/m}$, and instead of an energy or internal energy (pressure) equation, we have the constraint equation $T =$constant. Using $T =$ constant and $I =$ constant we eliminate the pressure and find that

$$\frac{1}{2u^2}\frac{du^2}{dr}\left[u^2 - \frac{2kT}{m}\right] = \frac{4kT}{mr} - \frac{GM}{r^2} \qquad (3)$$

The r.h.s. vanishes at the "critical radius"

$$r_c = \frac{GMm}{4kT} \qquad (4)$$

Between the coronal base at $R = r_0$ and the critical radius, the right hand side of Eq(3) is negative, implying that the thermal energy per proton is less than the binding energy of the gravitational potential. At the critical radius the left hand side must also vanish. This implies that either $u = (2kT/m)^{1/2}$ or that $du^2/dr = 0$ at $r = r_c$. These conditions at $r_c$ give rise to the well known X-type topology of the solutions in the $u - r$ plane, which is illustrated in Fig (1). Different families of solutions behave differntly near the critical point, and one of Parker's contributions was to recognize how nature selects the supersonic solution.

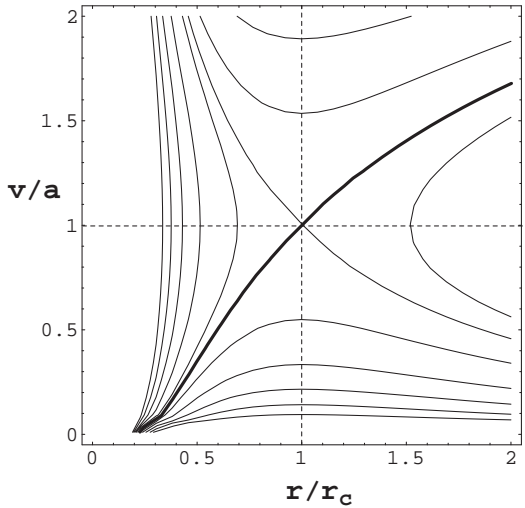

Figure 1: Solution topology for the isothermal model. Several types of solutions exist. The darker line is the Parker-type isothermal wind, which becomes supersonic at the critical point and is the only one the family of solutions that is consistent with the low pressure interstellar medium at the distant boundary. Courtesy of S. Owocki.

**Sonic Point in More Complete Models**

When additional effects are added to the wind equations, the mathematical nature of the equations changes, and the significance of the "critical point" may change or even become obscure. The "singular" aspect of the critical point in the isothermal model is in part due to the (over-)simplicity of the physics that it captures.[1] We can view the existence of the ambiguity of the solutions at the critical point as a carry-over from the time dependent case to the steady state case, reflecting how the characteristics of the time-dependent PDEs change when the flow speed exceeds the sound wave speed.

Physically, if the sound wave is the only (or, the fastest) mode of propagation of information within the medium, then the information in the supersonic region cannot make its way back to the subsonic region against the flow. The waves stagnate at the critical point, but can go no further. This separates the effects of the outer boundary conditions, and makes the problem really split in two, with the ends matched at the critical point, where the inner subsonic conditions influence the outer problem, but not vice versa.

These regions are also characterized by their different dominant balance of gravitational energy and thermal energy inside, when $r < r_c$, and the dominant balance of inertia and thermal pressure outside, for $r > r_c$. So the real importance of the critical point is twofold: it separates regions that have (1) differing energy balances, and (2), the differing regions of influence and causality associated with propagating wave modes. It is the existence of a boundary, whether sharp or diffuse, bewteen these regions, that has physical import.

Now suppose you add additional physics, such as magnetic field or heat conduction. The nature of the underlying PDEs changes (as do the type of characteristics), waves can become damped, and the speed of propagation of information may now exceed the sound speed. This would be the case, if, for example, the Alfvén speed is large, or of the electron thermal speed exceeds the proton thermal speed. In the corona both of these conditions are expected. Now information can readily propagate against the supersonic flow. This allows for other kinds of solar wind models, in which heat conduction and magnetic effects may be crucial (and there is plenty on this in the literature). With the addition of other effects such as

---

[1] As an example of the effects of oversimplification, consider a nonisothermal ideal gas equation of state such as an adiabatic relation $P \sim \rho^\gamma$, ratio of specific heats $\gamma$. The sound speed is $c_s^2 = \gamma P/\rho$, the adiabatic sound speed, and the critical point occurs when $u = c_s$, and *not* when $u = a$ as in Eq. (3).

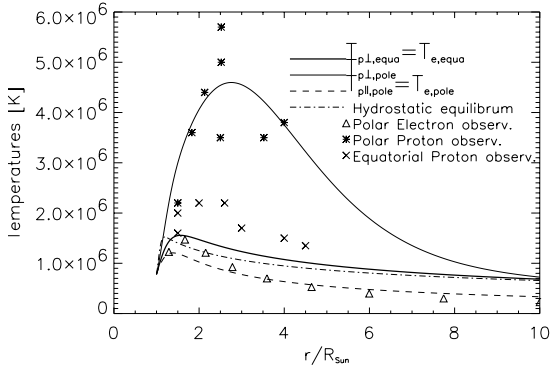

Figure 2: Radial profiles of temperatures from several observations, along with a model result, from Vasquez et al, 2004. Because the high latitude wind is much hotter, the sonic critical point is expect to be much lower in polar regions than in equatorial regions, at solar minimum.

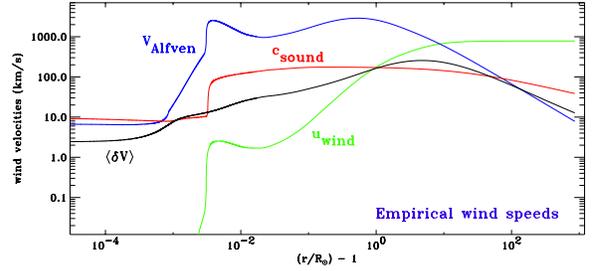

Figure 3: Radial profiles of wind speed, sound speed and Alfven speed from a coronal model, assembled by S. Cranmer. Note that the sonic critical point is around 1.5 $R_s$ and the Alfvén critical point is around 10 $R_s$. Courtesy of S. Cranmer.

heat conduction, the family of solutions can depart from the X-type topology seen in Fig (1), e.g., acquiring nodal, saddle, or multiple critical points, etc. (e.g., [16])

In spite of all this mathematical structure, a real time dependent solar wind, (or, a solar wind numerical model,) will either settle into some steady state, or thrash about in a non steady state, while always respecting imposed boundary and external conditions. In this sense the unattainable solutions discovered through critical point analysis are seen to be completely irrelevant. To a great degree the "critical point structure" of the equations is not reflected in the physics. Nevertheless, the existence of separate regions, implied by the critical point, having different energy balance and wave mode properties, is of great physical importance, e.g., in the Solar Probe mission. For example, it is widely accepted [10] that energy addition below the sonic point tends to increase the mass flux but not the wind speed, while energy deposition above the critical point accelerates the wind to higher speed without affecting mass flux, even in models that are not isothermal.

### Sonic Point in Polar Solar Wind

The basic physics of the rapid acceleration of the solar wind in the open field line coronal hole regions involves deposition of heat at low altitudes. Historically such models [12, 7] have included *ad hoc* heat functions, which serve to demonstrate the feasibility of such acceleration, which was thought to lead, possibly to a supersonic wind within several solar radii of the photosphere. It emerged as somewhat of a surprise when spectroscopic [8] and IPS [6] observations suggested that the speed exceeded hundreds of km/s by 2 $R_s$ or so. This was confirmed by UVCS spectroscopy [9], and associated with this it was determined that the proton temperature exceed several million $K$ within 1.5 $R_s$ or so of the photosphere. This important result, illustrated in Fig(2), has so far held up to additional careful reanalysis e.g., [4].

The sonic point generally moves downward for a hotter corona, as can be seen from the basic physics embodied in the isothermal model, Eq. (4), as $r_c \sim 1/T$. So in models with enhanced heating the sonic point is lower. In the Habbal et al (1995) model, an *ad hoc* heating function can give $r_c \approx 2.3 R_s$. Other models give similar results. Cranmer [3] has assembled coronal hole model based on launching of kink mode waves in the granulation regions, exciting Alfvén waves that propagate into the corona and induce heating through a turbulent cascade. It has been suggested (see Dmitruk et al, 2002) that a low frequency cascade of this type can supply sufficient energy to hear the corona, if the incident wave flux at the coronal base is adequate. Cranmer's model, which suggest that this is possible, gives radial profiles of the several relevant speeds which are shown in Fig (3). For more on this model, see [3]. Note that the sonic critical point in Cranmer's polar expanding flux tube model occurs near 2 $R_s$.

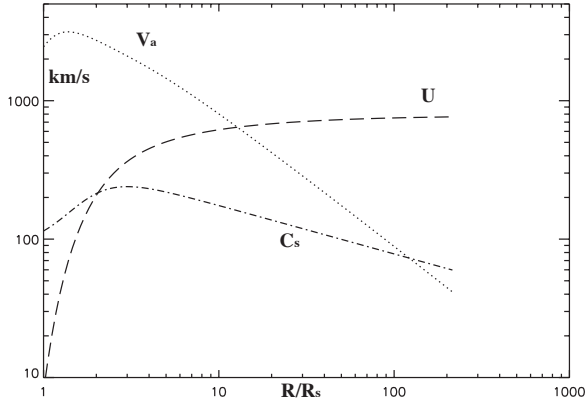

Figure 4: Radial profiles of wind speed, sound speed and Alfven speed from a coronal model, assembled by M. Velli. Note that the sonic critical point is around 2.5 $R_s$ and the Alfvén critical point is around 15 $R_s$. Figure courtesy of M. Velli.

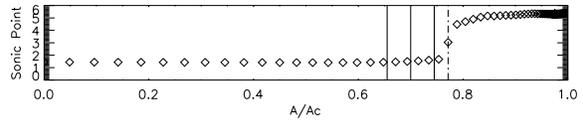

Figure 5: Using an observationally constrained model for coronal parameters, Vasquez et al, ApJ 2004 estimate this profile for the sonic point as a function of "field line label." The rapid change in sonic point near $A/A_c = 0.77$ corresponds to a field line latitude of about $65^o$ at the solar surface. At larger distances this field line is at the boundary of the streamer belt.

Another polar solar wind model [2] adopts a different approach, representing plume-like density enhancements in total pressure balance in the presence of Alfven waves and a prescribed temperature profile. This model also rapidly accelerates the wind and produces a sonic point around $r = 2.5R_s$. The results of this model are illustrated in Fig (4).

**Latitude dependence of the sonic point**

More elaborate models of coronal structure, including latitude dependence, have been based on semi-empirical approaches, for example by Sittler and Guhathakurta [13] and Vasquez et al [14]. Here we discuss only the latter model, which employs observational constraints, e.g., on temperature profiles, and force balance, to deduce further properties, such as an outflow velocity map and other diagnostics that can be visualized in a meridional plane. From these Vasquez et al find the latitudinal dependence of the sonic critical point. This is shown here in Fig(5), where it can be seen that the critical radius remains near 1.5 $R_s$ in the polar regions down to the field line that defines the streamer belt boundary. At that point it jumps rather sharply to 4 $R_s$ and then increases to about 5.5 $R_s$ at the equator.

It should be emphasized that even the most complete models at present do not incorporate transient effects at the scale of several hours or less, which are expected in both open and closed field line regions. All sorts of time dependent phenomena - footpoint motion, plumes, reconnection at various scales, turbulence, etc. may induce time dependence to the coronal structure, and variation in the base conditions such as temperature. This almost certainly would have significant impact on the sonic point, which as a consequence may be found at a variable radial distance. This radial variability has been emphasized by Coles (private communication) and others. Evidence for variability in speed at a given radius (integrated over line-of-sight) comes from IPS observations such as those of Grall et al [6]. See also [11, 15]. Fig.(6) shows a well known plot of the Grall et al observations, which demonstrates both the rapid

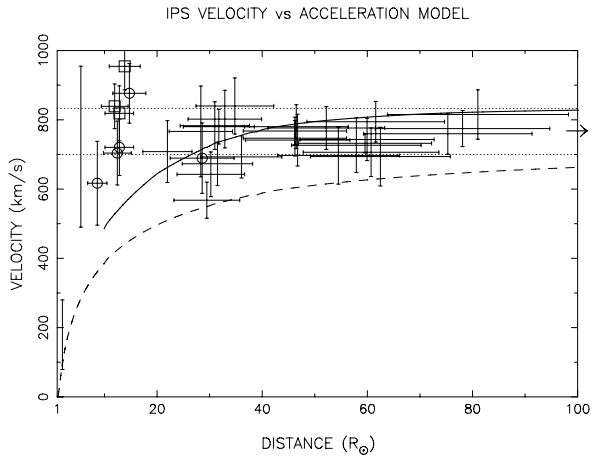

Figure 6: Observations of polar coronal outflow velocity from IPS and Spartan data, compared with some models, from Grall et al (1996), showing rapid acceleration of the solar wind. The location of the Alfvénic point may be in the range 10-20 $R_s$.

acceleration of the wind as well as the variability of the observed speeds.

**Sonic Point and Solar Probe**

To a great degree the mathematical notion of critical points, will have little or no impact on the Solar Probe mission. However there remains some physical interest in understanding at what radius the solar wind flow becomes supersonic, and the above summary concludes that the place that this happens around 1.5 to perhaps 3 $R_s$ at high latitudes and 5 to 8 $R_s$ in the streamer belt. These positions are expected to be unsteady and may not even define a very smooth surface. It is probably better to think of the transonic *region,* rather than the sonic *point.*

There will be some practical consequences, from the point of view of instrumentation, to planning for when the spacecraft will be passing through a subsonic medium. One well-discussed effect is the biasing of measured particle distribution due to the peculiar speed associated with the sum of space craft and solar wind speeds.

The estimated subsonic region and spacecraft trajectory can be easily visualized in an approximate way. In Fig (7) a 4 $R_s$ perihelion trajectory of Solar Probe is shown superposed on a coronagraph image (this is the cover of a prior SDT report). To illustrate a possible subsonic region, a $2R_s$ by 5 $R_s$ shaded ellipse is centered on the sun. A subsonic solar wind will be seen at most for ±1 hour from perihelion. If the subsonic regions is confined to the streamer belt this time may be only ±20 minutes or so.

A related issue is the Mach number of the spacecraft speed. Spacecraft speed is around 300 km/s at the equatorial perihelion and about 200 km/s at the polar passes at around $10R_s$. Therefore it appears likely that the spacecraft will be moderately supersonic in the plasma frame throughout the primary mission. [2]

From a theoretical point of view the volume of space on either side of the sonic point (whether it is sharply defied or not), may be quite interesting, especially if due to transient activity the mission sometimes sees this transition in open field line regions. An essential goal of Solar probe is to understand the energy flows and energy budgets that enter in to the origin and acceleration of the wind. Comparing subsonic to supersonic regions may be a fruitful activity in this regard, as would be an examination of any additional or peculiar turbulence that occurs in this region.

**Alfvén critical point**

In spite of the amount written about coronal magnetic fields, there are no reliable determinations of the magnetic field strength throughout most of the regions that Solar probe will pass through. Therefore even though densities are relatively well known, Alfvén speeds are not. This means that, for example, in open field line regions, for which the Alfvén speed is often quoted as 1000-2000 KM/s between 2 and 10 $R_s$, it is difficult to rule out the possibility that, in some subvolumes, the Alfvén speed may be an order or magnitude smaller than this. This is a large uncertainty.

---

[2]The s/c spaeed can add to, or oppose, the wind speed, and on the outbound orbit one might wonder if this can drop the speed relative to the plasma to subsonic values. However, even though the s/c reaches a maximum radial speed of about 150 km/s in the primary mission, this occurs at high latitude where the wind speed should be in excess of 500 km/s by 6-9 $R_s$. The sound speed is probably between 150 and 250 km/s for the entire primary mission (outside the streamer belt.

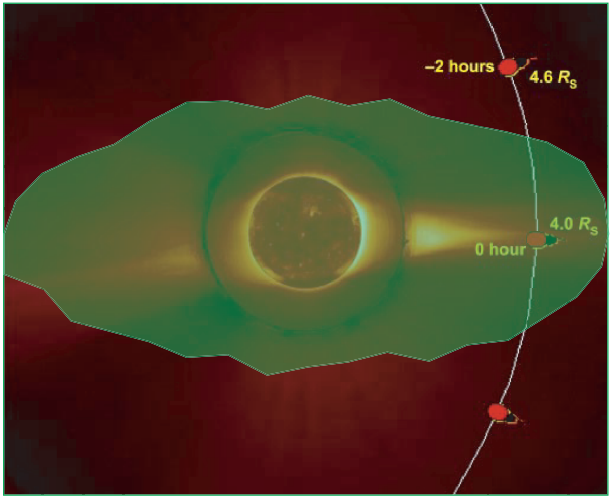

Figure 7: Coronagraph images with a Solar Probe trajectory superposed (from prior SDT Report cover).) A sun-centered shaded elliptical region of overall dimension approximately 5 x 2 Rs was added to denote the approximate region of expected subsonic flow.

In contrast, we think that we know quite a bit about the topology of magnetic fields from high resolution imaging by TRACE, LASCO, etc., but even these interpretations are rarely 100 % reliable, because they are typicaly looking at light intensity, which goes like density time temperature to some power, and these may be dynamic quantities. Most of our beliefs about Alfvén speeds come from using modeling to extrapolate the stronger photospheric fields, which can be measured.

With this in mind, many models make predictions about the Alfvén critical point, where solar wind speed $U = $ Alfvén speed $V_a$. Older estimates for the equatorial Alfvén point fall in the range of $10-20R_s$. Some years ago there was a controversial Russian IPS paper that claimed to show enhanced turbulence (scatter) near 15-20 $R_s$ and this was interpreted as indicating the Alfvén point. (See also [11, 15].) From more modern models, one tends to see numbers such as those in Figures (3 and 4) above, from the Cranmer and Velli et al models, respectively. These show the Alfvén point at around 10 and 15 $R_s$ respectively. Sittler and Guhathakurta (1999) cite $13R_s$ from their semi-empirical model. These seem to denote a consensus range for the coronal hole Alfvén point.

Like the sonic point, the Alfvén critical point does not have a strict significance except in certain limited models, e.g., a model in which the Alfvén speed is the fastest speed of propagation of information and energy that is available. However this is not the case in the kinetic solar wind plasma, which allows electron motion, heat conduction and a large number of kinetic wave modes that can supply super-Alfvénic communication. Nevertheless, much of the fluid-scale energy in the fluctuating solar wind is expected to be in low compressibility, transverse fluctuations, akin to Alfvén modes. So for these, the Alfvén point signifies their subsequent inability to communicate upstream, in analogy to the disconnection that occurs for sound waves outside the sonic point. In both cases there is a possibility that stagnation of the inward wave causes a buildup of fluctuations that might enhance turbulent interactions in that region.

Most likely the entire primary Solar Probe Mission will be inside the Alfvén point if "primary" is defined as $\pm 10 R_s$. For scientific value, it would be very useful to explore the inner super-Alfvénic wind, since this is new territory. Having the instruments fully operational in this region would also guarantee that the region around the Alfvén point would be observed in detail as well. The nature of the plasma and its fluctuations in this region may give important information about what kind of fluctuations and waves are emanating from lower altitudes.

### Conclusion

Neither sonic nor Alfvénic points are expected to be important in the sense of "critical points," since this characterization reflects in part the oversimplification in elementary models. However both "regions" of transition may be interesting with regard to exami-nation of energy budgets, especially in fluctuations. They may also have some bearing on optimization or operation of some instruments. Minimally, detection and characterization of these transition regions will constrain models of the dynamic and steady corona and solar wind.

**Aknowledgements**